\documentclass[
  reprint,
  superscriptaddress,
  amsmath,amssymb,
  aps,
  pra,
  floatfix,
  twocolumn
]{revtex4-2}

\usepackage[utf8]{inputenc}
\usepackage[dvipsnames]{xcolor}
\usepackage{graphicx}
\usepackage{enumitem}
\usepackage{siunitx}
\DeclareSIUnit\gauss{G}
\usepackage{booktabs}
\usepackage{multirow}
\usepackage{mathtools}
\usepackage{flushend}
\usepackage[switch]{lineno}
\usepackage{soul}
\usepackage{braket}
\usepackage{times}
\usepackage{bm}

\newcommand{\probP}{\text{I\kern-0.15em P}}

\makeatletter 
\renewcommand{\fnum@figure}{\textbf{Fig.~\thefigure}}
\makeatother

\begin{document} 

\title{Optical Creation of Synthetic Microgravity for Quantum Degenerate Gases}

\author{Catie LeDesma}
\thanks{These authors contributed equally to this work.}

\author{Kendall Mehling}
\thanks{These authors contributed equally to this work.}

\author{Tristan Rojo}

\author{Murray Holland}
\email{murray.holland@colorado.edu}
\affiliation{JILA \& Department of Physics, University of Colorado Boulder, Boulder, CO, 80309-0440}

\date{\today}

\begin{abstract}
Microgravity environments provide unique opportunities for ultracold-atom experiments by enabling long interrogation times and reduced acceleration-induced dynamics. However, their realization has largely been restricted to specialized facilities such as drop towers, sounding rockets, and space-based laboratories. Here we realize synthetic microgravity for quantum degenerate gases using optically engineered force landscapes that compensate Earth's gravity to the milli-g level while maintaining continuous confinement of the atomic ensemble. These force landscapes are generated by dynamically painted optical dipole potentials and calibrated in situ through Bloch oscillations in a vertical optical lattice, enabling precise control of the residual acceleration. We use this capability to demonstrate matter-wave beam splitting with arm separations of several hundred microns. We further implement a Bloch-band atom interferometer in which interaction-induced dephasing is strongly suppressed through controlled three-dimensional expansion in the synthetic microgravity potential. This reduction of mean-field effects restores near-$\sqrt{N}$ scaling of interferometric sensitivity for large quantum degenerate ensembles. Our results establish a versatile platform for realizing synthetic microgravity with trapped quantum gases in terrestrial laboratories, bringing the advantages of microgravity experiments to continuously operating systems and opening new opportunities for quantum sensing, matter-wave interferometry, and precision measurements.
\end{abstract}

\maketitle 

On Earth, gravity imposes fundamental limitations on ultracold-atom experiments, restricting both achievable interrogation times and the properties of trapped quantum gases~\cite{Becker2018, Elliott2018}. In free-fall atom interferometers, atoms accelerate at $9.8~\mathrm{m/s^2}$ and rapidly leave the interaction region, typically limiting interrogation times to below one second in laboratory-scale systems unless large vertical baselines are employed \cite{PhysRevLett.66.2693, PhysRevLett.67.181}. Gravity also induces sag in trapped atomic ensembles, distorting confinement potentials and complicating the realization of homogeneous many-body systems \cite{Hung2011, Gaunt2013, Navon2021}. These limitations have motivated the development of a diverse range of microgravity platforms, including drop towers, parabolic-flight campaigns, sounding rockets, Einstein elevators, and space-based laboratories~\cite{canuel2018exploring, zhan2020zaiga, asenbaum2020atom, badurina2020aion, abe2021matter, Barrett2016, lachmann2021ultracold, Pelluet2025, Chinese_Space_Gyro, zhang2026inorbittestweakequivalence, williams2024pathfinder}. These facilities have enabled major advances in atom interferometry, quantum sensing, and ultracold-atom physics, including Bose--Einstein condensation and matter-wave interferometry in space-based environments \cite{Becker2018, Chinese_Space_Gyro, williams2024pathfinder}. However, they require specialized infrastructure, provide only intermittent access, and are difficult to translate into continuously operating or deployable quantum devices \cite{Barrett2016, Becker2018, Elliott2018, Bongs2019, Pelluet2025}.

The need for laboratory-based microgravity is particularly acute in trapped and guided matter-wave systems. Unlike free-fall interferometers, these systems can in principle support long interrogation times within compact geometries~\cite{Barrett2016, Bongs2019}, but gravity severely restricts the large-volume expansion needed to reach low-density regimes. Consequently, atomic densities remain high during interrogation, enhancing interaction-induced dephasing, reducing coherence times, and degrading interferometric sensitivity \cite{Robins_confined, Gupta_meanfield, Alauze_2018}. Accessing dilute conditions typically requires either reducing atom number or allowing extended ballistic expansion, both of which compromise signal strength and experimental flexibility \cite{Robins_magGrad, hough2026}. As a result, trapped matter-wave systems face a fundamental trade-off between maintaining confinement and achieving the low densities required for long coherence times and high-precision measurements.

In this work, we pursue an alternative route toward realizing synthetic microgravity without large-scale free-fall infrastructure. Rather than engineering the external environment to approximate free fall, we engineer the force landscape experienced by the atoms using dynamically painted optical dipole potentials \cite{Ryu2013PaintedPotential, Grimm2000OpticalDipole, Gaunt2013BoxTrap}. By programming a spatially tailored vertical potential gradient, we generate an optical buoyancy force that counteracts gravity while preserving continuous confinement of the atomic ensemble. This approach enables precise control of the residual acceleration experienced by the atoms and provides access to microgravity-like conditions within a compact terrestrial apparatus.

\begin{figure*}[t!]
\centering
\includegraphics[width=0.7\textwidth]{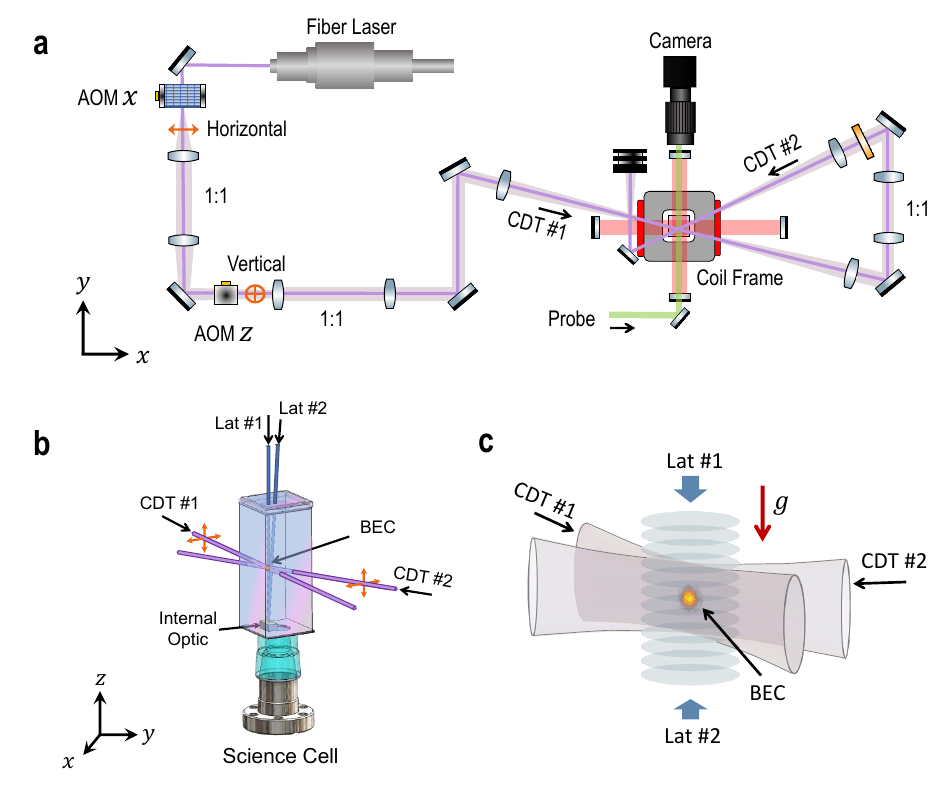}
\caption{\textbf{Experimental apparatus and trapping geometry.} (\textbf{a}) Experimental setup. A high-power optical dipole beam (purple) is generated by a fiber laser and passes through two acousto-optic modulators (AOMs) that provide beam steering along the horizontal ($x$) and vertical~($z$) directions. The beam is focused through the science chamber and recycled to form an elongated crossed-dipole trap (CDT). Magneto-optical trap (MOT) beams are shown in red, and a probe beam (green) is used for absorption imaging. (\textbf{b}) Illustration of the science cell and trapping geometry. The CDT beams (purple) intersect within the science chamber, while an internal optic enables implementation of a vertical optical lattice (blue) without requiring optical access from below. (\textbf{c}) Representation of the synthetic microgravity potential. Time-averaged beam painting generates a highly elongated optical dipole potential that overlaps with the vertical optical lattice. By programming a spatially varying optical potential along the direction of gravity, an optical buoyancy force is produced that compensates gravitational acceleration while maintaining continuous confinement of the Bose--Einstein condensate located at the intersection of the CDT and lattice beams.
}
\label{fig:apparatus}
\end{figure*}

A key challenge in realizing synthetic microgravity through engineered force landscapes is the precise calibration and verification of the residual force experienced by the atoms. Because many of the anticipated benefits arise when the effective acceleration is reduced by several orders of magnitude below Earth's gravitational field, even small systematic offsets must be accurately characterized and controlled. In this work, we address this challenge using Bloch oscillations of a Bose--Einstein condensate in a vertical optical lattice \cite{PhysRevLett.76.4508}, providing us with a sensitive in situ probe of the net force acting on the atoms. This approach enables calibration at the milli-g level and establishes quantitative control over the synthetic microgravity environment.

To independently verify the suppression of residual forces, we demonstrate a machine-learned matter-wave beam-splitter sequence based on modulation of the lattice phase and track the resulting atomic trajectories through absorption imaging. Fits to the measured wave-packet motion yield accelerations consistent with zero within experimental uncertainty, providing direct evidence for a near-force-free environment. We then demonstrate the utility of this platform for trapped atom interferometry. By enabling controlled three-dimensional pre-expansion of the atomic ensemble while maintaining confinement, the engineered force landscape suppresses interaction-induced dephasing and substantially reduces contrast loss arising from mean-field effects. This mitigation restores interferometric sensitivity toward the expected $\sqrt{N}$ scaling \cite{HL_Holland}, where $N$ is the particle number. Together, these results establish synthetic microgravity as a powerful new tool for quantum-gas experiments, compact matter-wave interferometry, and precision quantum sensing.

\begin{figure*}[t!]
    \centering
    \includegraphics[width=1.00\textwidth]{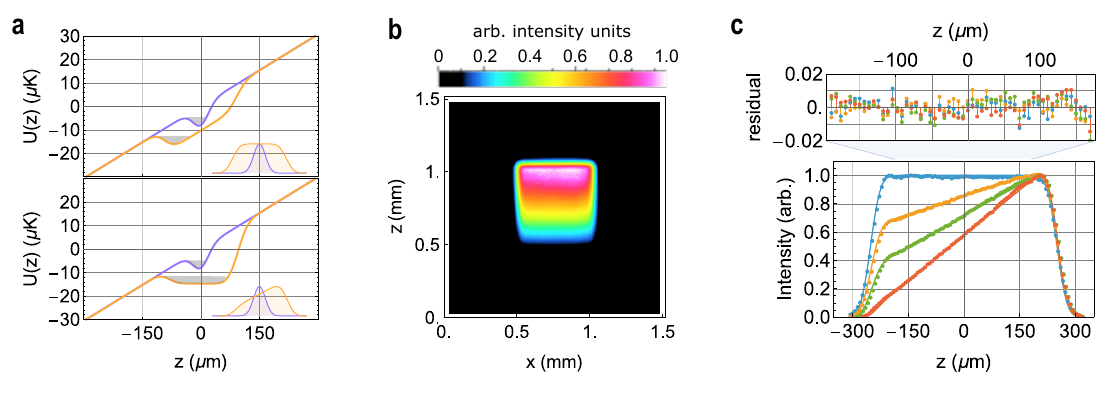}
    \caption{\textbf{Gravity compensation through engineered intensity gradients.} (\textbf{a}) Calculated trapping potentials along the vertical axis under gravity. Upper: Comparison of a static dipole trap (purple) and a painted trap scanned over $\pm100~\mu$m (orange) with no applied intensity gradient. Despite the increased trap volume, gravity localizes atoms near the lower region of the potential, indicated by the shaded gray areas. Insets show the corresponding normalized intensity profiles. Lower: Trapping potentials with an applied intensity gradient. The painted trap~(orange) is scanned over $\pm100~\mu$m with a gradient of approximately $78\%$, generating an upward optical force that compensates gravity and restores access to the full trap volume. Insets again show the corresponding intensity profiles. (\textbf{b}) Experimentally measured intensity image of a two-dimensional painted potential acquired with a beam profiler. The potential is generated by scanning over $\pm300~\mu$m in both the $x$ and $z$ directions, corresponding to an effective trap size of approximately $600~\mu$m $\times$ $600~\mu$m. A $100\%$ intensity gradient is applied along the vertical ($z$) direction. (\textbf{c}) Horizontally integrated intensity profiles for applied gradients of $0\%$ (blue), $25\%$ (yellow), $50\%$ (green), and $70\%$~(red). Inset shows residuals from linear fits demonstrating the measured fidelity of the programmed intensity gradients.}
    \label{fig:compensation}
\end{figure*}

\vspace*{1pc}
\noindent{\large\bf Results}\newline
\noindent{\bf Engineering neutral buoyancy into optical potentials} \newline
Painted optical potentials provide a flexible and compact method for engineering trapping geometries and force landscapes for ultracold atoms \cite{Ryu2013PaintedPotential, Henderson2009AOM}. In this approach, a tightly focused, far-detuned laser beam is rapidly scanned using an acousto-optic modulator (AOM), producing a time-averaged optical potential when the scan frequency exceeds the characteristic atomic timescale~\cite{Grimm2000OpticalDipole}. By controlling the position, amplitude, and dwell time of the scanned beam, a wide range of trapping geometries can be realized, including box potentials, ring traps, and waveguides \cite{Ryu2013PaintedPotential, Gaunt2013BoxTrap, Henderson2009AOM, Roy_2016}. 

Our experimental apparatus is based on the platform described in Refs.~\cite{LeDesmaVectorSciAdv, LeDesmaGateset,LeDesma2024Thesis}, with modifications enabling two-dimensional beam painting for dynamic control of the trapping geometry and force landscape. The layout (Fig.~\ref{fig:apparatus}a) uses two orthogonally-oriented AOMs to independently control the horizontal ($x$) and vertical~($z$) position of an optical crossed dipole trap (CDT). After a first pass through the vacuum cell, the trapping beam is recycled with orthogonal polarization to produce a second beam that inherits the same painting dynamics, forming a fully programmable three-dimensional trapping geometry. A separate vertically oriented optical lattice is overlapped with the condensate (Fig.~\ref{fig:apparatus}b,c), providing a sensitive probe of the effective acceleration and a platform for Bloch-band atom interferometry. In order to realize a time-averaged potential in this platform, the raster frequency must exceed the trapping and vibrational frequencies of the CDT and lattice ($10^2$--$10^4$~Hz range). 

Although painted potentials enable large and highly tunable trapping geometries, they do not by themselves mitigate the effects of gravity. Increasing the geometric extent of a trap does not increase the volume accessible to the atoms, because gravitational sag shifts the equilibrium position toward the bottom of the potential. As a result, atoms can occupy only a small fraction of the available trap volume, leading to elevated densities and preventing access to the dilute regimes needed for long coherent evolution. The upper panel of Fig.~\ref{fig:compensation}a illustrates this effect, showing that even a substantially expanded painted trap remains populated only near the gravitational minimum.

To overcome this limitation, we engineer the spatial intensity profile of the painted dipole trap to generate a compensating optical force. The total potential experienced by the atoms is
\begin{equation}
U(\mathbf{r},t)=U_\mathrm{D}(\mathbf{r},t)+mgz,
\end{equation}
where $U_\mathrm{D}(\mathbf{r},t)$ is the painted optical dipole potential at position $\mathbf{r}=(x,y,z)$ and time $t$, and the term $mgz$ accounts for gravity, with $m$ the atomic mass and $g$ the gravitational acceleration at the Earth's surface. Because the optical dipole force is proportional to the gradient of the trapping potential, introducing a controlled vertical intensity gradient generates an upward force that partially or completely compensates gravity. When the optical and gravitational forces balance, the atoms reach a condition analogous to neutral buoyancy, experiencing an effective near-weightless environment while remaining continuously confined within the trap. The resulting force-balanced potential restores access to the full trap volume, as illustrated in the lower panel of Fig.~\ref{fig:compensation}a.

Generating the synthetic microgravity potential in multiple dimensions requires the trapping beam to be dynamically rastered in both the horizontal and vertical directions. The beam is scanned linearly back and forth along the $x$ and $z$ axes, producing a time-averaged intensity distribution given by the convolution of the scan trajectory with the underlying Gaussian beam profile of waist~$w_0$. Gravity compensation is achieved by applying a position-dependent amplitude weighting during the scan, thereby imprinting a controlled linear intensity gradient along the vertical direction. No corresponding gradient is applied along the horizontal axis because gravity acts only along $z$.

To verify that the programmed light gradients are faithfully realized, we directly measure the time-averaged beam profiles using a beam profiler. The exposure time is chosen to be much longer than the scan period, ensuring that the recorded image reflects the cycle-averaged intensity distribution. A representative two-dimensional beam profile is shown in Fig.~\ref{fig:compensation}b. The vertical intensity distribution is obtained by integrating the image along the horizontal direction, and examples for several programmed gradients are shown in Fig.~\ref{fig:compensation}c. Linear fits to the central region of each profile are used to extract the corresponding intensity gradients. The measured gradients agree well with the programmed values, with residual deviations dominated by imaging noise.

\vspace*{1pc}
\noindent{\bf In situ calibration via Bloch oscillations} \newline
To accurately quantify the degree of gravitational compensation, we use the atoms themselves as an direct probe of the residual force. We accomplish this using the vertically oriented optical lattice introduced in Fig.~\ref{fig:apparatus}, which serves both as a sensitive probe of residual acceleration and, later, as a platform for atom interferometry. In the presence of the lattice, the total potential includes the dipole and gravitational contributions discussed above together with the lattice potential
\begin{equation}
U_\mathrm{L}(z,t)=\frac{V_0}{2}\cos\bigl(2kz+\phi(t)\bigr),
\end{equation}
where the wavenumber $k=2\pi/\lambda$ is determined by the lattice wavelength $\lambda$, and $\phi(t)$ is the lattice phase. The lattice depth we use is $V_0=10E_\mathrm{R}$, where $E_\mathrm{R}=\hbar^2k^2/(2m)$ is the recoil energy.

This approach is particularly important because small imperfections in the programmed intensity gradient can produce appreciable deviations in the realized force landscape. When atoms are loaded into the optical lattice, a constant force along the lattice axis induces Bloch oscillations \cite{PhysRevLett.76.4508}. The oscillation frequency $f_\mathrm{B}$ is directly proportional to the applied force,
\begin{equation}
f_\mathrm{B}=\frac{Fd}{2\pi\hbar},
\end{equation}
where $F$ is the net force acting on the atoms and $d=\lambda/2$ is the lattice spacing. Bloch oscillations therefore provide a sensitive and direct measure of the residual acceleration.

Bloch oscillations have long been established as a powerful tool for precision force measurements \cite{PhysRevLett.106.038501, PhysRevLett.92.230402, Cadoret_2008}. Originally introduced in the context of electron dynamics in periodic solids \cite{1929ZPhy...52..555B,1934RSPSA.145..523Z}, they arise generically for particles in a spatially periodic potential subjected to a constant force. Ultracold atoms in optical lattices provide a particularly clean realization of this phenomenon, with eigenstates organized into energy bands indexed by a band number $n$ and a continuous quasimomentum $q$ restricted to the first Brillouin zone,
\begin{equation}
q\in\Bigl[-\frac{\pi\hbar}{d},\frac{\pi\hbar}{d}\Bigr].
\end{equation}
In the absence of an external force, atoms prepared in the lowest band ($n=0$) remain in stationary Bloch states. When a force is applied, the quasimomentum evolves according to $\dot q = F$, causing the atoms to traverse the Brillouin zone. Upon reaching the zone edge, Bragg reflection returns the atoms to the opposite side of the band, resulting in periodic Bloch oscillations with period $\tau_\mathrm{B}=1/f_\mathrm{B}$.

\begin{figure*}[t!]
    \centering
    \includegraphics[width=0.70\textwidth]{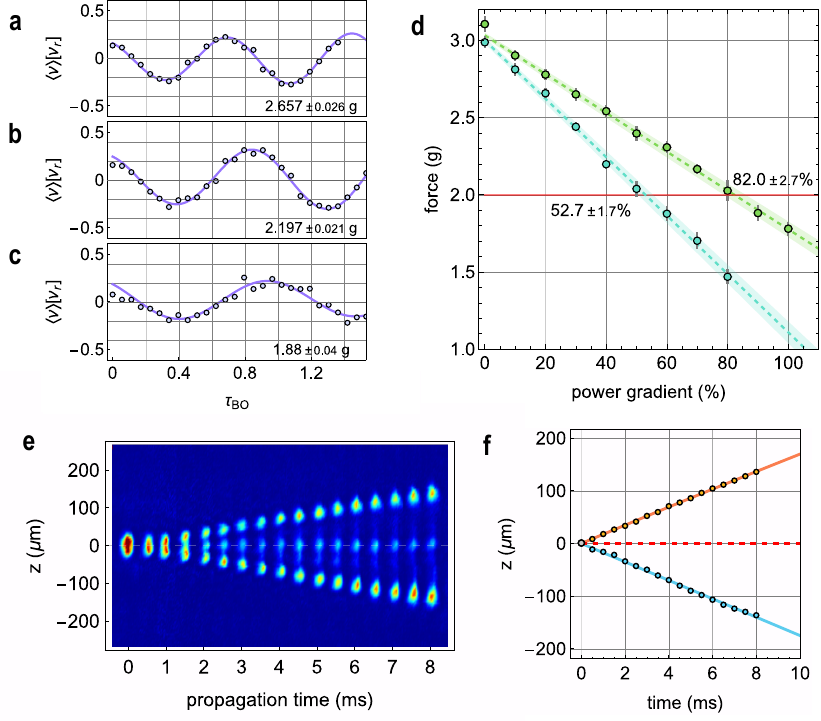}
    \caption{\textbf{Bloch-oscillation calibration of synthetic microgravity and matter-wave beam splitting.} (\textbf{a--c}) Representative Bloch oscillation measurements for applied intensity gradients of 20\%, 40\%, and 60\%, yielding calibrated forces of $2.657 \pm 0.026$\,g, $2.197 \pm 0.021$\,g, and $1.88 \pm 0.04$\,g, respectively. (\textbf{d}) Forces extracted from Bloch oscillation frequencies as a function of applied intensity gradient for a $\pm200$\,\textmu m trap (aqua) and a $\pm300$\,\textmu m trap (green); dashed lines are linear fits. The measurements from \textbf{a--c} correspond to three of the points on the $\pm200$\,\textmu m trap calibration curve. The red horizontal line marks the condition at which the magnitude of the optical force equals gravity, intersecting the $\pm200$\,\textmu m and $\pm300$\,\textmu m calibration curves at gradients of $52.7 \pm 1.7$\% and $82.0 \pm 2.7$\%, respectively. (\textbf{e}) In situ images showing the spatial evolution of the Bose-Einstein condensate as a function of propagation time in a $\pm$200\,\textmu m gravity-corrected painted potential following application of a beamsplitter protocol. (\textbf{f}) Extracted trajectories of the two beamsplitter arms from \textbf{e} (orange and blue), with linear fits showing relative motion of each arm post-splitting and the atoms separating at $\pm4\hbar k$. The dashed red line indicates $z = 0$.}
    \label{fig:bo}
\end{figure*}

For these measurements, condensates containing on the order of $2\times10^4$ rubidium-87 atoms were prepared in a nearly pure Bose--Einstein condensate and adiabatically loaded into the lowest energy band of the vertical optical lattice, providing a low-entropy, narrow-momentum probe suitable for precision force measurements. Following lattice loading, the painted potential was abruptly switched on, exposing the atoms to the engineered force landscape while preserving their initial quasimomentum distribution. The subsequent Bloch dynamics were allowed to evolve for several milliseconds before being measured using time-of-flight absorption imaging, from which the quasimomentum evolution was extracted. By repeating this procedure for programmed intensity gradients ranging from $0\%$ to $100\%$, we directly measured the force landscape experienced by the atoms as a function of the applied optical gradient.

In the absence of an optical gradient, atoms adiabatically loaded into the vertical lattice undergo Bloch oscillations at a frequency determined by gravity. As the optical force approaches gravitational compensation, the net force acting on the atoms is reduced and the Bloch frequency correspondingly decreases. In the region near perfect compensation, however, the oscillation frequency becomes sub-Hz, making precise measurements challenging because of finite interrogation times and experimental noise. To maintain high measurement sensitivity throughout the calibration procedure, we deliberately introduce a known offset force by accelerating the optical lattice upward at $2g$ by a relative frequency sweep of the vertical lattice lasers. In the moving frame with the lattice, atoms confined in a flat painted potential then experience a net force of $3g$ due to the equivalence principle. As the gradient is increased, the compensating optical force reduces the net acceleration and the Bloch frequency decreases accordingly. Complete gravitational compensation corresponds to a net residual force of~$2g$, providing a convenient and experimentally accessible reference point for calibration.

This offset-biasing strategy produces a readily measurable linear relationship between Bloch frequency and applied intensity gradient (Fig.~\ref{fig:bo}a--d), enabling direct calibration of the engineered force landscape in units of gravitational acceleration. Representative Bloch oscillation measurements for several applied gradients are shown in Fig.~\ref{fig:bo}a--c, while Fig.~\ref{fig:bo}d summarizes the extracted forces over the full gradient range. Linear fits identify the gradient required for gravitational compensation and determine the corresponding residual acceleration with an uncertainty of approximately $3\times10^{-2}g$. We emphasize that this uncertainty reflects the calibration accuracy of the static force landscape rather than the ultimate metrological sensitivity of the system. For atom interferometry and related sensing applications, sensitivity is instead governed by interferometer geometry, force fluctuations, and phase noise during a given measurement sequence.

\begin{figure*}[t!]
    \centering
    \includegraphics[width=1.00
\textwidth]{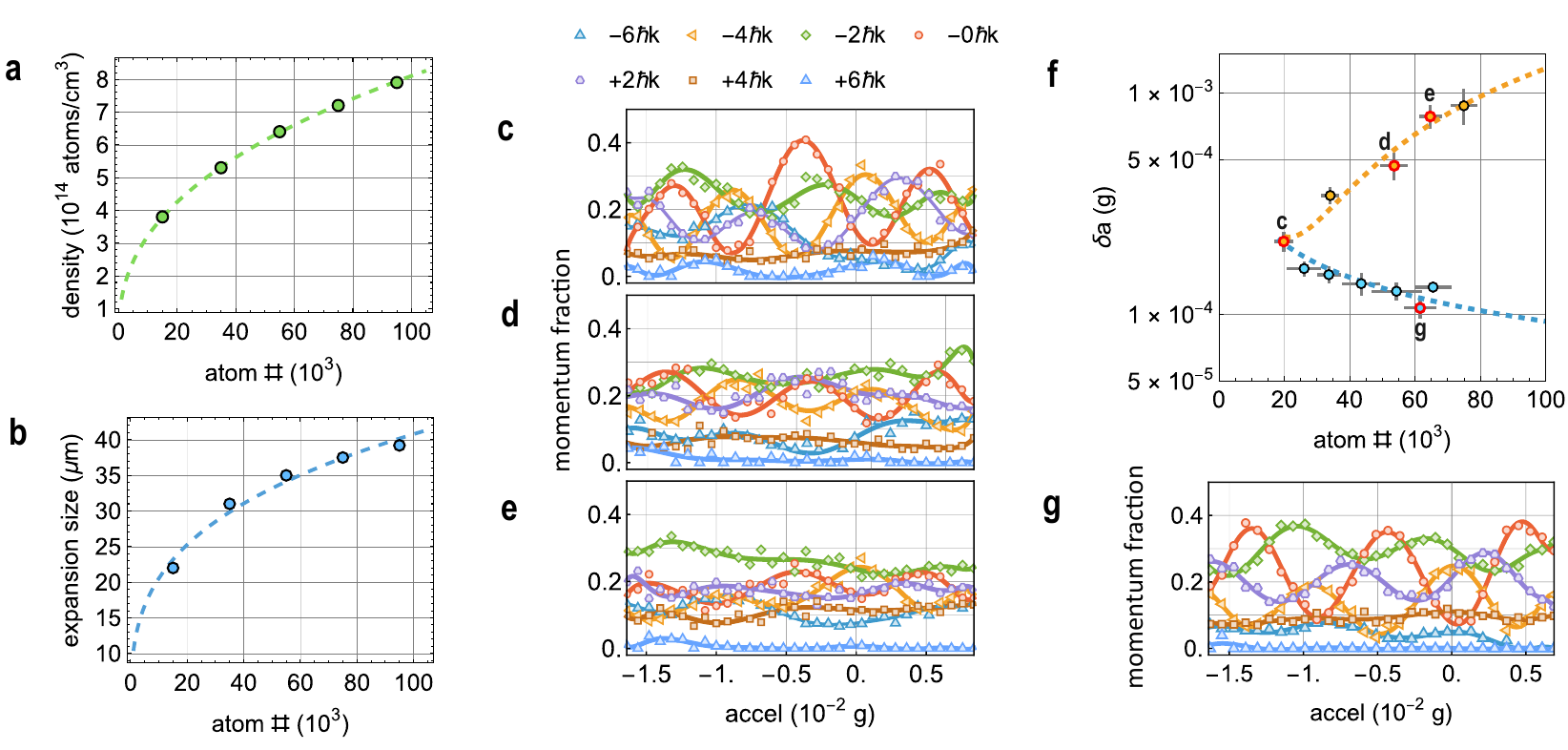}
    \caption{\textbf{Interaction-induced dephasing and its suppression via pre-expansion.}
(\textbf{a}) Peak density scaling from mean-field theory as a function of atom number for a select number of points (solid circles) in the unexpanded crossed dipole trap, following the expected $n_0 \propto N^{2/5}$ Thomas--Fermi scaling (dashed line). 
(\textbf{b}) Required trap expansion distances calculated from mean-field theory to maintain a target peak density of $n_0 < 2.75 \times 10^{14}$\,cm$^{-3}$ as a function of atom number for selected values (solid circles), with a dashed line fit to guide the eye.  (\textbf{c--e}) Interference fringes (momentum fraction vs.\ applied acceleration) at the atom numbers indicated by labels \textbf{c--e} in \textbf{f}, measured without pre-expansion and showing progressive loss of fringe contrast with increasing density. Colors indicate momentum components from $-6\hbar k$ to $+6\hbar k$. (\textbf{f}) Acceleration sensitivity $\delta a$ as a function of atom number without (yellow) and with (blue) pre-expansion in the 3D painted potential. Labeled points mark the atom numbers corresponding to the fringe plots in \textbf{c--e} and \textbf{g}. Without expansion, sensitivity degrades with atom number, deviating significantly from the ideal $1/\sqrt{N}$ scaling (blue dashed line, referenced to the lowest-density case); with pre-expansion, sensitivities recover toward this scaling. The yellow dashed curve is an empirical fit included solely to guide the eye. (\textbf{g}) Representative high-contrast interference fringe at large atom number following pre-expansion, demonstrating strong suppression of interaction-induced dephasing and subsequent full revival of interferometer contrast.}
    \label{fig:fringe}
\end{figure*}

\vspace*{1pc}
\noindent{\bf Matter-wave dynamics in synthetic microgravity}\newline
While Bloch oscillations provide a sensitive calibration of the residual force, they do not directly reveal how atoms propagate through the engineered potential landscape. To independently verify the realization of a synthetic microgravity environment, we implement a coherent matter-wave beam-splitting sequence and directly track the resulting atomic trajectories. Following calibration of the force gradient, the crossed dipole trap is painted over a vertical range of $\pm200~\mu$m and operated at the gradient corresponding to gravitational compensation, producing a large synthetic microgravity region spanning approximately $400~\mu$m. After loading the atoms into the optical lattice, they are coherently split using a machine-learned $\pm4\hbar k$ beam-splitter protocol (see Methods) and are allowed to evolve, creating two coherent wave packets that dynamically separate. 

An in situ absorption-image sequence showing the propagation of the separating matter waves over 8~ms is presented in Fig.~\ref{fig:bo}e. The two wave packets move apart by more than $300~\mu$m while remaining confined within the engineered potential, directly demonstrating full access to the enlarged force-compensated trap volume. The corresponding trajectories extracted from multi-Gaussian fits to the absorption images are shown in Fig.~\ref{fig:bo}f. Both arms exhibit uniform motion over the full observation time, indicating that the atoms experience negligible residual acceleration while traversing the synthetic microgravity region.

To quantify this behavior, we fit each trajectory to the center-of-mass kinematic form
\begin{equation}
x(t)=vt+\frac{1}{2}at^2.
\end{equation}
For the upper arm, we obtain $v_1=(-17.2\pm0.1)\,\mu\mathrm{m/ms}$ and $a_1=(-0.14\pm0.22)\,\mu\mathrm{m/ms^2}$, while for the lower arm we find $v_2=(17.2\pm0.1)\,\mu\mathrm{m/ms}$ and $a_2=(-0.045\pm0.140)\,\mu\mathrm{m/ms^2}$. The extracted relative velocity is
\begin{equation}
v_{\mathrm{rel}}=(34.44\pm0.08)\,\mu\mathrm{m/ms}
=(7.98\pm0.02)\,\hbar k/m,
\end{equation}
in excellent agreement with the expected $8\hbar k$ relative momentum predicted for the machine-learned  beamsplitter sequence. Averaging the fitted accelerations yields
\begin{equation}
\bar{a}=(-0.007\pm0.012)\,g,
\end{equation}
which is statistically consistent with zero. These matter-wave dynamics therefore provide an independent verification of the Bloch-oscillation calibration method and confirm that atoms propagate through the engineered potential with no measurable residual force. The uncertainty is dominated by the imaging resolution (5.45~$\mu$m/pixel), which limits the precision of the extracted accelerations.

\vspace*{1pc}
\noindent{\bf Bloch-band atom interferometry and interaction effects}\newline
Bose-condensed atomic ensembles are widely used in precision atom interferometry due to their intrinsic coherence properties, including narrow momentum distributions and long spatial coherence lengths \cite{Hagley2001, Shin2004}. However, their high phase-space density also enhances mean-field interactions. For species with appreciable scattering lengths and no readily accessible Feshbach resonance, these interactions generate spatially dependent phase shifts, leading to dephasing between interferometer arms, reduced fringe visibility, and degraded sensitivity \cite{Shin2004, Cronin2009Review}. In the absence of interactions, interferometric sensitivity is expected to follow the standard quantum limit \cite{HL_Holland}, scaling as $1/\sqrt{N}$. In practice, however, mean-field dephasing often prevents straightforward access to this scaling at large atom numbers. This limitation is particularly restrictive because weakly interacting regimes are typically reached only in relatively small ($\,\lesssim 10^4$ atoms) or highly expanded condensates, making it difficult to simultaneously achieve high atom number and long coherence times. A central motivation for the synthetic microgravity platform developed here is that it enables substantial three-dimensional expansion while maintaining confinement. This reduces atomic density and suppresses interaction-induced dephasing, which directly impacts metrological applications.

A standard approach to mitigating interaction effects is to allow the condensate to expand prior to interferometry. During expansion, the atomic density and corresponding mean-field energy decrease rapidly, suppressing phase diffusion and improving coherence. Free-fall interferometers and microgravity platforms naturally provide access to long expansion times and therefore low-density regimes. In contrast, trapped interferometers maintain confinement throughout the interrogation sequence, preventing extended ballistic expansion and leaving the atomic density comparatively high. As a result, interaction-induced dephasing can become a dominant limitation on both interrogation time and interferometric sensitivity. Although decompression strategies can be implemented in trapped systems, gravity strongly constrains the achievable expansion geometries and durations in terrestrial laboratories. Here, we overcome this limitation using the synthetic microgravity environment created by the painted optical potential. By dynamically expanding the trapping geometry prior to interferometry, the condensate density can be substantially reduced while maintaining continuous confinement. Simultaneously, the engineered optical buoyancy force preserves gravitational compensation throughout the expansion, enabling access to low-density regimes without releasing the atoms or relying on extended free-fall times.

To determine the expansion required to suppress interaction-induced dephasing at large atom numbers, we first model how the density of the trapped condensate evolves with increasing atom number. Fig.~\ref{fig:fringe}a shows the peak density $n_0$ in the unexpanded CDT, which follows the expected Thomas-Fermi scaling in a harmonic oscillator, $n_0\propto N^{2/5}$. As a consequence, increasing atom number rapidly drives the system toward interaction-dominated regimes. To suppress mean-field dephasing we therefore chose to limit the peak density to $n_0<2.75\times10^{14}\,\mathrm{cm}^{-3}$. This value was determined empirically through experimental observation of full interferometer fringe contrast for comparable densities. In order to dilute condensates of increasing atom number, progressively larger trap volumes are required. Fig.~\ref{fig:fringe}b shows the corresponding expansion distances for a fixed trap depth of approximately~$3\,\mu\mathrm{K}$. For condensates containing more than a few times $10^4$ atoms, the required expansion lengths approach $\sim\!\!40\, \mu m$, placing them well within the regime enabled by the synthetic microgravity potentials developed here.

To directly quantify the impact of interactions on phase coherence, a phase-sensitive measurement of the accumulated relative quantum phase is required. We therefore implement a Bloch-band  interferometer \cite{ledesma2023machinedesignedopticallatticeatom,LeDesmaVectorSciAdv,LeDesmaGateset,MarcoMag} using Bose-Einstein condensed ensembles spanning a range of atom numbers and densities. Prior to interrogation, the condensate is decompressed in the painted potential according to the scaling shown in Fig.~\ref{fig:fringe}b. The interferometric sequence is then applied with typical interrogation times of 1~ms. Atom-optic components are implemented using machine-learned control sequences generated through precise modulation of the lattice phase (see Methods). In selected cases, additional expansion was applied to achieve complete revival of interferometer contrast.

Figs.~\ref{fig:fringe}c--e show representative interferometer fringes measured at increasing atom numbers without pre-expansion. In our Bloch-band interferometer, the output signal is obtained by monitoring the populations of seven momentum states spanning $[-6\hbar k,-4\hbar k,\ldots,6\hbar k]$ as a function of applied acceleration. The resulting oscillatory population response encodes the accumulated relative phase between the interferometer arms. 
The fringe visibility in Fig.~\ref{fig:fringe}c for this multi-port interferometer is consistent with full contrast as validated by theory. As the atom number and corresponding density increase, the fringe visibility progressively decreases due to enhanced mean-field dephasing. This loss of contrast directly degrades the precision with which the applied acceleration can be estimated. The corresponding acceleration sensitivities are shown in Fig.~\ref{fig:fringe}f (yellow), where increasing atom number leads to a pronounced departure from the ideal $1/\sqrt{N}$ scaling expected in the absence of interaction-induced dephasing (blue dashed line).

With pre-expansion, however, this behavior is reversed. As shown in Fig.~\ref{fig:fringe}f and the representative fringe data of Fig.~\ref{fig:fringe}g, expansion within the synthetic microgravity potential strongly suppresses interaction-induced dephasing and restores high-contrast interference even at large atom numbers. The corresponding acceleration sensitivities (Fig.~\ref{fig:fringe}f, blue) recover toward the expected $1/\sqrt{N}$ scaling, demonstrating that the dominant interaction-driven limitations have been substantially mitigated.

These results provide strong evidence that controlling interaction-induced dephasing is essential for realizing the intrinsic scaling advantages of Bose-Einstein condensed interferometers. More generally, programmable trapping geometries establish a route to low-density operation in confined systems without requiring extended free-fall, enabling compact quantum sensors that maintain long coherence times while operating at atom numbers that would otherwise be strongly limited by mean-field effects.

\vskip1pc
\noindent{\large\bf Discussion}\newline
In summary, we demonstrate a synthetic microgravity platform for quantum degenerate gases using programmable painted potentials. By engineering spatial intensity gradients, we compensate gravitational acceleration while maintaining continuous confinement. Using Bloch oscillations, we calibrate the effective force to the milli-g level and verify force cancellation via matter-wave propagation and beam-splitting dynamics. We further show that the same platform enables controlled decompression of Bose-condensed ensembles, reducing interaction-induced dephasing and restoring near-$\sqrt{N}$ interferometric scaling.

This approach differs from conventional reduced-gravity methods that rely on modifying the external environment or operating in free fall. Here, gravity is effectively compensated by tailoring the optical potential experienced by the atoms. The resulting microgravity condition arises from engineered force landscapes rather than mechanical isolation or ballistic motion. While not equivalent to true free-fall microgravity in all respects, it directly mitigates key limitations in ultracold atom experiments, including gravitational sag, restricted expansion, and finite interrogation volume. Importantly, continuous trapping is preserved, enabling compact and continuously operating implementations.

A key practical advantage of painted potentials is their simplicity and flexibility. They require only standard optical components (AOMs, electronics, and imaging optics) and programmable waveform generation (AWG or FPGA). Different trapping geometries are produced by updating radio-frequency (RF) waveforms rather than modifying experimental hardware. The repetition rate is primarily limited by Bose-Einstein condensate production times \cite{Bongs2019Review, MewesBEC}, and with demonstrated sub-Hz BEC production \cite{Hebert2020SubSecondErbium, PhysRevA.111.L061301}, this approach is compatible with  high-duty-cycle operation. Because the system is fully optically defined, the architecture is inherently compact and compatible with portable implementations.

Several technical limitations presently limit performance metrics. The achievable trap size is constrained by optical power, scan range, and acousto-optic modulator bandwidth, which together limit both spatial extent and confinement strength. In the present implementation, synthetic microgravity regions of approximately $400~\mu\mathrm{m}$ were realized using 1.6 W of average input optical power for the first pass of the CDT. Larger regions can be achieved either through increased optical power or by reducing the required confinement strength. These are technical rather than fundamental constraints, and improvements in optical power handling, beam shaping, and scanning bandwidth should enable larger, flatter force-balanced regions. Similar gains could be achieved by operating the dipole trap closer to atomic resonance, where the increased optical polarizability reduces the laser power required to generate a given trapping potential.

It is important to distinguish the static residual accelerations measured via Bloch oscillations from the sensitivity limit of an interferometric measurement. The former characterizes offsets in the engineered force landscape and therefore only appears as a bias in the measured interferometric fringe. On the other hand, sensitivity is determined by the enclosed space-time area of the interferometer together with force fluctuations and phase noise within the measurement bandwidth. Consequently, the milli-g-level calibration uncertainty reported here should not be interpreted as a fundamental sensitivity limit. Rather, it quantifies the accuracy with which the static force landscape is realized, while the ultimate measurement sensitivity depends on both the interferometer geometry and the stability of the engineered potential.

An additional advantage of the force-engineering approach is that the compensated acceleration need not be limited to gravity. In practical deployments, ultracold-atom systems operating aboard moving vehicles or mobile platforms may experience substantial inertial accelerations arising from vehicle motion, vibration, or changing orientation. Because the synthetic microgravity environment is generated through programmable optical force landscapes, the same framework can in principle be extended to compensate or bias against these accelerations as well. This capability suggests a route toward dynamically reconfigurable trapping and interferometric architectures in which the local acceleration environment is actively engineered to match the operational conditions of the sensor.

In broader perspective, this work demonstrates that microgravity can be treated as an engineered force landscape rather than solely as an environmental condition achieved through free fall or operation in specialized facilities. By directly programming the forces experienced by the atoms, key features of reduced-gravity operation can be reproduced within a continuously operating laboratory system. From this perspective, synthetic microgravity represents a complementary paradigm to conventional reduced-gravity platforms, in which the local acceleration environment becomes a tunable experimental parameter rather than a fixed property of the laboratory.

\vskip1pc
\noindent{\large\bf Methods}\newline
\noindent{\bf Experimental apparatus}\newline
Bose-Einstein condensates of $^{87}$Rb are produced in a custom double-MOT vacuum system. Following laser cooling and trapping in a standard six-beam magneto-optical trap and sub-Doppler cooling stage, approximately $5\times10^6$ atoms are loaded into a crossed optical dipole trap (CDT). For more details of the rapid all-optical BEC production see Refs.~\cite{LeDesmaVectorSciAdv, LeDesmaGateset,LeDesma2024Thesis}. 

The CDT is generated using a 1064~nm, 50~W fiber laser and consists of two intersecting dipole beams formed by recycling a single beam through the science cell. The beams intersect at an angle of approximately $32^\circ$ and have waists of approximately $40~\mu$m at the atom position. The recycled beam is orthogonally polarized relative to the first pass to suppress interference effects. A small fraction of the beam is sampled before entering the science cell and monitored with a photodiode, enabling active feedback stabilization of the trap intensity at the sub-percent level over experimental timescales. Forced evaporative cooling is performed through a sequence of linear power ramps over approximately 5~s, producing Bose--Einstein condensates containing up to $1\times10^5$ atoms with total experimental cycle times of approximately 10~s.

The vertical optical lattice is formed from two free-space counter-propagating 1064~nm laser beams derived from a separate 50~W fiber laser. The beams are reflected from an internal mirror within the vacuum system, enabling lattice implementation despite limited optical access below the science cell. Each lattice beam passes through an independent AOM, allowing dynamic control of the lattice phase $\phi(t)$ through controlled frequency detuning between the counter-propagating beams. This capability is used both to apply controlled inertial biases during force calibration and to implement atom-optic operations such as beam splitting and reflection. As with the CDT, the intensity of each lattice beam is actively stabilized using photodiode-based feedback derived from light sampled before entering the science cell. The lattice beam waist at the atom position is approximately $250~\mu$m, and the lattice depth used throughout this work is typically $V_0 = 10E_{\mathrm{R}}$.

Two-dimensional beam painting is implemented using two orthogonally-oriented AOMs, which independently control beam deflection along the horizontal ($x$) and vertical ($z$) axes. Each AOM is driven by an independent RF channel generated by a dual-channel AWG with 1~ns timing resolution. RF carrier frequencies are chosen near the AOM center frequencies, and controlled modulation of the drive signals produces angular beam deflections that are converted into spatial displacements through a sequence of 1:1 imaging telescopes. The calibrated displacement at the atom position is approximately $50~\mu$m/MHz for both painting axes. Because the CDT is formed by recycling the trapping beam through the science cell, an additional 1:1 telescope is inserted between passes to re-image the painted beam profile and compensate the image inversion that would otherwise occur. This ensures that the programmed intensity gradients and painted trapping geometries maintain the same orientation for both dipole-trap passes.

Because the diffraction efficiency of each AOM varies across its RF sweep range, uncorrected frequency scans produce spatially nonuniform intensity distributions. Such distortions can introduce unintended variations in trap depth and force gradients, degrading the fidelity of the programmed potential landscape. To compensate for this effect, the RF amplitude applied to each AOM is pre-corrected using independently measured diffraction-efficiency curves. The correction is implemented point-by-point during waveform generation, ensuring that the realized intensity profile accurately reproduces the programmed trap geometry and optical force landscape.

\vskip1pc
\noindent{\bf Painting details}\newline
Painted potentials are generated by rapidly rastering the focused dipole beam using the AOM-driven steering system. For scan frequencies that significantly exceed the characteristic trap frequencies ($\sim10^2$--$10^3$~Hz), the atoms experience a time-averaged potential determined by the convolution of the scan trajectory with the underlying Gaussian beam profile. In this regime, both the trapping geometry and the effective force landscape can be programmed by controlling the spatial trajectory and intensity weighting of the scanned beam.

For a beam scanned uniformly over an interval $[-d,d]$, the normalized one-dimensional time-averaged intensity envelope can be calculated analytically from the convolution of the scan trajectory with the Gaussian beam profile,
\begin{equation}
E(\beta,d)=
\frac{
\mathrm{Erf}\!\left[\sqrt{2}(d-\beta)/w_0\right]
+
\mathrm{Erf}\!\left[\sqrt{2}(d+\beta)/w_0\right]
}{
2\,\mathrm{Erf}\!\left[\sqrt{2}d/w_0\right]
},
\end{equation}
where $\beta\!\in\!\{x,z\}$ denotes the spatial coordinate along a painted direction, $w_0$ is the beam waist, and $\mathrm{Erf}$ denotes the error function.

For independent and sufficiently rapid scans along orthogonal axes, the full two-dimensional envelope is well approximated as separable,
\begin{equation}
E(x,z,d_x,d_z)\approx E(x,d_x)\,E(z,d_z).
\end{equation}
Separability enables independent control of the horizontal and vertical trapping dimensions, Along the unpainted dimension~($y$), the potential envelope is determined solely by the native Gaussian beam profile.

A controlled vertical intensity gradient is imposed by applying an additional amplitude weighting during the raster cycle. Because gravity acts only along the vertical direction, no corresponding gradient is applied along the horizontal axis. The imposed weighting function is
\begin{equation}
G(z,\alpha,d_z)=
(1-\alpha)+\frac{\alpha}{2d_z}(z+d_z),
\end{equation}
where $\alpha=0$ corresponds to a uniform intensity profile and $\alpha=1$ corresponds to a $100\%$ linear intensity ramp across the painted vertical extent. Multiplication of the painted intensity envelope by $G(z,\alpha,d_z)$ generates the optical force responsible for gravitational compensation, with the parameter~$\alpha$ controlling the strength of the engineered force gradient. To ensure that changes in $\alpha$ maintain a constant peak RF power, the programmed intensity profile is renormalized prior to waveform generation.

The synthetic microgravity potential is realized by combining the painted intensity envelope with the programmed vertical amplitude weighting. The resulting painted dipole potential is
\begin{equation}
U_{\mathrm{D}}(x,z)=
U_0\,E(x,z,d_x,d_z)\,G(z,\alpha,d_z),
\end{equation}
where $U_0$ sets the overall trap depth. Including the optical lattice and gravity, the total time-averaged potential experienced by the atoms is
\begin{equation}
U(x,z,t)=U_{\mathrm{D}}(x,z)+U_\mathrm{L}(z,t)+mgz.
\end{equation}

The horizontal and vertical scan frequencies are chosen to be incommensurate (typically 77~kHz and 100~kHz, respectively) to ensure uniform filling of the painted volume and to avoid closed scan trajectories. These frequencies are substantially larger than the trap frequencies while avoiding parametric heating resonances of the optical lattice system. They are also selected to lie within the effective modulation bandwidth of the AOMs, which is approximately 120~kHz for the scan amplitudes employed in this work. Operating within this regime preserves stable beam steering and minimizes distortions of the programmed intensity profile during rapid rastering. Once calibrated, the resulting painted force landscapes remain stable over timescales of several weeks without requiring recalibration.

\vskip1pc
\noindent{\bf Bloch-band interferometry}\newline
Bloch-band interferometry is implemented through precise control of the vertical optical lattice. In this architecture, the lattice serves simultaneously as the trapping potential, inertial reference frame, and atom-optic element. Beam-splitter, reflection, and recombination operations are realized through controlled modulation of the lattice phase, allowing coherent manipulation of atomic momentum states within the lattice band structure. The required phase-control waveforms are synthesized using an AWG and applied through controlled modulation of the lattice AOM drive frequencies. For the lattice depth used in this work momentum populations are resolved across seven diffraction orders spanning $[-6\hbar k,\ldots,6\hbar k]$, which constitute the measured output channels of the interferometer.

For studies of interaction-induced dephasing, the condensate is first adiabatically decompressed within the painted optical potential by slowly increasing the trap dimensions over approximately 200~ms. This expansion reduces the atomic density while maintaining continuous confinement. Once the desired trap size is reached, the optical lattice is ramped on over a comparable timescale to ensure adiabatic loading into the lattice band structure despite the relatively weak lattice confinement. To isolate the effects of density reduction from those of gravitational compensation, the painted potential is subsequently turned off prior to the interferometric sequence. Interferometry is then performed in a freely falling lattice frame, realized by chirping the lattice frequency to match the gravitational acceleration $g$. In this configuration, gravity is effectively canceled during interrogation while the lattice confinement remains unchanged, allowing the influence of condensate expansion on mean-field dephasing to be studied independently.

The control protocols used to implement beam-splitter, reflection, and recombination operations were generated using a reinforcement-learning optimization framework developed in previous work \cite{RLMatter_Chih}. The resulting lattice-phase sequences are optimized for high-fidelity coherent manipulation of the atomic momentum states while maintaining robustness to experimental imperfections. Typical interrogation times used throughout this work were 1~ms.

As in Ref.~\cite{LeDesmaVectorSciAdv}, interferometric sensitivity is quantified using the classical Fisher information (CFI) together with the associated Cramér--Rao bound. For a given probability distribution over the seven output momentum modes, the CFI determines the maximum information obtainable per measurement about the applied acceleration. Because the CFI is intrinsically a single-particle metric, estimating the total experimental sensitivity additionally requires knowledge of the effective number of independent trials contributing to the measured signal, denoted $N_{\mathrm{trial}}$.

Experimentally, the CFI was obtained by averaging over five independent scans of applied acceleration. To determine $N_{\mathrm{trial}}$, we acquired 100 independent experimental realizations at an operating point corresponding to large Fisher information ($-0.01 g$). Assuming multinomial counting statistics across the interferometer output modes, we fit a multinomial model to randomly sampled subsets of the measured data with varying sample sizes and extracted an effective trial number from the resulting scaling behavior. The reported interferometric sensitivities were then calculated from the measured CFI and the extracted value of $N_{\mathrm{trial}}$. 

Because the interferometer is implemented using a free-space counter-propagating optical lattice, relative phase fluctuations between the lattice beams are directly imprinted onto the atomic phase and therefore contribute noise to the interferometric signal. In this sense, the optical lattice serves as the local oscillator of the interferometer and requires active phase stabilization even for the relatively short interrogation times employed here. To suppress both low- and high-frequency phase noise, the lattice phase is actively stabilized using the phase-locking architecture described in Ref.~\cite{MehlingPhase}. This stabilization significantly reduces technical phase noise and ensures that the measured sensitivities are not limited by lattice coherence.

\vskip1pc
\noindent{\large\bf Acknowledgments}\newline
We thank Michelle Stevens and Penina Axelrad for helpful discussions. This work was supported by a Space Technology Research Institutes grant from NASA's Space Technology Research Grants Program (Grant No.\ 80NSSC23K1343). Support is acknowledged from NSF Grant Nos.\ 2317149 and 2016244, and by the U.S. Department of Energy, Office of Science, National Quantum Information Science Research Center and Quantum Systems Accelerator. We acknowledge support from DARPA Grant No.\ HR0011-25-3-E02.  
\bibliography{ref}

\end{document}